\documentclass[12pt]{article}
\begin{document}
\title{Fuzzy Space Time, Quantum Geometry and Cosmology}
\author{B.G. Sidharth\\
Centre for Applicable Mathematics \& Computer Sciences\\
Adarsh Nagar, Hyderabad - 500 063, India}
\date{}
\maketitle
\begin{abstract}
It is argued that a noncommutative geometry of spacetime leads to
a reconciliation of electromagnetism and gravitation while
providing an underpinning to Weyl's geometry. It also leads to a
cosmology consistent with observation. A few other ramifications
are also examined.
\end{abstract}
\section{Introduction}
The Theory of Relativity (Special and General) and Quantum Theory
have been often described as the two pillars of twentieth century
physics. Yet it was almost as if Rudyard Kipling's "The twain
shall never meet" was true for these two intellectual
achievements. For decades there have been fruitless attempts to
unify electromagnetism and gravitation, or Quantum Theory and
General Relativity. As Wheeler put it \cite{MWT}, the problem has
been, how to incorporate curvature into Quantum Theory or spin
half into General Relativity. At the same time it is also
remarkable that both these disparate theories share one common
platform: An underlying differentiable space time manifold, be it
the Reimannian spacetime of General Relativity or the Minkowski
spacetime of Relativistic Quantum Theory (including Quantum Field
Theory). However this underlying common feature has been
questioned by Quantum Gravity on the one hand and Quantum
SuperStrings on the other, which try to unify these two branches
(Cf.ref.\cite{bgs} and several references therein). We will now
argue that unification and a geometrical structure for Quantum
Theory are possible if differentiable spacetime is discarded in
favour of fuzzy spacetime. At the same time this has many
ramifications and leads to a cosmology which is consistent with
the latest iconoclastic observations, for example that the
universe is accelerating and expanding for ever while the fine
structure constant seems to be changing with time.
\section{Quantum Geometry}
One of the earliest attempts to unify electromagnetism and
gravitation, was Weyl's gauge invariant geometry. The basic idea
was \cite{weyl} that while
\begin{equation}
ds^2 = g_{\mu \nu} dx^\mu dx^\nu\label{e1}
\end{equation}
was invariant under arbitrary transformations in General
Relativity, a further invariant, namely,
\begin{equation}
\Phi_\mu dx^\mu\label{e2}
\end{equation}
which is a linear form should be introduced. $g_{\mu \nu}$ in
(\ref{e1}) would represent the gravitational potential, and
$\Phi_\mu$ of (\ref{e2}) would represent the electromagnetic field
potential. As Weyl observed, "The world is a $3 + 1$ dimensional
metrical manifold; all physical field - phenomena are expressions
of the metrics of the world. (Whereas the old view was that the
four-dimensional metrical continuum is the scene of physical
phenomena; the physical essentialities themselves are, however,
things that exist "in" this world, and we must accept them in type
and number in the form in which experience gives us cognition of
them: nothing further is to be "comprehended" of
them.)$\cdots$"\\
This was a bold step, because it implied the relativity of
magnitude multiplied effectively on all components of the metric
tensor $g_{\mu \nu}$ by an arbitrary function of the coordinates.
However, the unification was illusive because the $g_{\mu \nu}$
and $\Phi_\mu$ were
really independent elements\cite{Einstein}.\\
A more modern treatment is recapitulated
below \cite{berg}.\\
The above arbitrary multiplying factor is normalised and we
require that,
\begin{equation}
| g_{\mu \nu} | = -1,\label{e3}
\end{equation}
For the invariance of (\ref{e3}), $g_{\mu \nu}$ transforms now as
a tensor density of weight minus half, rather than as a tensor in
the usual theory. The covariant derivative now needs to be
redefined as
\begin{equation}
T^{\iota \cdots}_{\kappa \cdots ,\sigma} = T^{\iota
\cdots}_{\kappa \cdots ,\sigma} + \Gamma ^\iota_{\rho \sigma}
T^{\rho \cdots}_{\kappa \cdots} - \Gamma^\rho_{\kappa \sigma}
T^{\iota \cdots}_{\rho \cdots} - n T^{\iota \cdots} _{\kappa
\cdots} \Phi_\sigma ,\label{e4}
\end{equation}
In (\ref{e4}) we have introduced the $\Phi_\mu$, and $n$ is the
weight of the tensor density. This finally leads to
(Cf.ref.\cite{berg} for details).
\begin{equation}
\Phi_\sigma = \Gamma^\rho_{\rho \sigma},\label{e5}
\end{equation}
$\Phi_\mu$ in (\ref{e5}) is identified with the electromagnetic
potential, while $g_{\mu \nu}$ gives the gravitational potential
as in the usual theory. The affine connection is now given by
\begin{equation}
\Gamma^\lambda_{\iota \kappa} = \frac{1}{2} g^{\lambda \sigma}
(g_{\iota \sigma ,\kappa} +  g_{\kappa \sigma ,\iota} - g_{\iota
\kappa ,\sigma}) + \frac{1}{4} g^{\lambda \sigma} (g_{\iota
\sigma}\Phi_\kappa + g_{\kappa \sigma} \Phi_\iota - g_{\iota
\kappa} \Phi_\sigma ) \equiv \left(\begin{array}{ll} \lambda \\
\iota \kappa
\end{array}\right)\label{e6}
\end{equation}
The essential point, and this was the original criticism of
Einstein and others, is that in (\ref{e6}), $g_{\mu \nu}$ and
$\Phi_\mu$ are independent entities.\\
Let us now analyze the above from a different perspective. Let us
write the product $dx^\mu dx^\nu$ of (\ref{e1}) as a sum of half
its anti-symmetric part and half the symmetric part. The invariant
line element in (\ref{e1}) now becomes $(h_{\mu \nu} + \hbar_{\mu
\nu}) dx^\mu dx^\nu$ where $h$ and $\hbar$ denote the
anti-symmetric and symmetric parts respectively of $g$. $h$ would
vanish unless the commutator
\begin{equation}
[dx^\mu , dx^\nu ] \approx l^2 \ne 0\label{e7}
\end{equation}
$l$ being some fundamental minimum length. In this case, under
reflection, $h_{\mu \nu} \to - h_{\mu \nu}$ as in the case of the
tensor density metric tensor above.\\
To proceed further, we observe that the noncommutative geometry
given in (\ref{e7}) was studied by Snyder and others. In this case
it has been shown in detail by the author \cite{bgscsf,fuzzy} that
under an infinitesimal Lorentz transformation of the wave
function,
\begin{equation}
| \psi' > = U(R)| \psi >\label{e8}
\end{equation}
we get
\begin{equation}
\psi' (x_j) = [1 + \imath \epsilon (\imath \epsilon l_{jk}x_k
\frac{\partial} {\partial x_j}) + 0 (\epsilon^2)] \psi
(x_j)\label{e9}
\end{equation}
Equation (\ref{e9}) has been shown to lead to the Dirac equation
when $l$ is the Compton wavelength. Indeed, Dirac himself had
noted that his electron equation needed an average over spacetime
intervals of the order of the Compton scale to remove
zitterbewegung effects and give meaningful physics. This again has
recently been shown to be symptomatic of an underlying fuzzy
spacetime described by a noncommutative space time geometry
(\ref{e7}) \cite{esn}.\\
The point here is that under equation (\ref{e7}), the coordinates
$x^\mu \to \gamma^{(\mu)} x^{(\mu)}$ where the brackets with the
superscript denote the fact that there is no summation over the
indices.  Infact, in the theory of the Dirac equation it is well
known \cite{bade}.that,
\begin{equation}
\gamma^k \gamma^l + \gamma^l \gamma^k = - 2g^{kl}I\label{e10}
\end{equation}
where $\gamma$'s satisfy the usual Clifford algebra of the Dirac
matrices, and can be represented by
\begin{equation}
\gamma^k = \sqrt{2} \left(\begin{array}{ll}
0 \quad \sigma^k \\
\sigma^{k*} \quad 0
\end{array}\right)\label{e11}
\end{equation}
where $\sigma$'s are the Pauli matrices. As noted by Bade and
Jehle (Cf.ref.\cite{bade}), we could take the $\sigma$'s or
$\gamma$'s in (\ref{e10}) and (\ref{e11}) as the components of a
contravariant world vector, or equivalently one could take them to
be fixed matrices, and to maintain covariance, to attribute new
transformation properties to the wave function, which now becomes
a spinor (or bi-spinor). This latter has been the traditional
route, because of which the Dirac wave function has its
bi-spinorial character. In this latter case, the coordinates
retain their usual commutative character. It is only when we
consider the equivalent former alternative, that we return to the
noncommutative
geometry (\ref{e7}).\\
That is in the usual commutative spacetime the Dirac spinorial
wave functions conceal the noncommutative
character (\ref{e7}).\\
Indeed we can verify all these considerations in a simple way as
follows:\\
First let us consider the usual space time. This time the Dirac
wave function is given by
$$\psi = \left(\begin{array}{ll}
\chi \\ \Theta
\end{array}\right),$$
where $\chi$ and $\Theta$ are spinors. It is well known that under
reflection while the so called positive energy spinor $\Theta$
behaves normally, $\chi \to -\chi , \chi$ being the so called
negative energy spinor which comes into play at the Compton scale
\cite{bd}. Because of this property as shown in detail
\cite{fuzzy}, there is now a covariant derivative given by, in
units, $\hbar = c=1$,
\begin{equation}
\frac{\partial \chi}{\partial x^\mu} \to [\frac{\partial}{\partial
x^\mu} - n A^\mu]\chi\label{e12}
\end{equation}
where
\begin{equation}
A^\mu = \Gamma^{\mu \sigma}_{\sigma} = \frac{\partial} {\partial
x^\mu} log (\sqrt{|g|)}\label{e13}
\end{equation}
$\Gamma$ denoting the Christofell symbols.\\
$A^\mu$ in (\ref{e13})is now identified with the electromagnetic
potential, exactly as in
Weyl's theory.\\
What all this means is that the so called ad hoc feature in Weyl's
unification theory is really symptomatic of the underlying
noncommutative space time geometry (\ref{e7}). Given (\ref{e7}) we
get both gravitation and electromagnetism in a unified picture.\\
Let us now consider the above ideas in the context of the
deBroglie-Bohm formulation \cite{cu}. We start with the
Schrodinger equation
\begin{equation}
\imath \hbar \frac{\partial \psi}{\partial t} = -
\frac{\hbar^2}{2m} \nabla^2 \psi + V \psi\label{e1a}
\end{equation}
In (\ref{e1a}), the substitution
\begin{equation}
\psi = Re^{\imath S/\hbar}\label{e2a}
\end{equation}
where $R$ and $S$ are real functions of $\vec r$ and $t$, leads
to,
\begin{equation}
\frac{\partial \rho}{\partial t} + \vec \nabla \cdot (\rho \vec v
) = 0\label{e3a}
\end{equation}
\begin{equation}
\frac{1}{\hbar} \frac{\partial S}{\partial t} + \frac{1}{2m} (\vec
\nabla S)^2 + \frac{V}{\hbar^2} - \frac{1}{2m} \frac{\nabla^2
R}{R} = 0\label{e4a}
\end{equation}
where
$$\rho = R^2 , \vec v = \frac{\hbar}{m} \vec \nabla S$$
and
\begin{equation}
Q \equiv - \frac{\hbar^2}{2m} (\nabla^2 R/R)\label{e5a}
\end{equation}
Using the theory of fluid flow, it is well known that (\ref{e3a})
and (\ref{e4a}) lead to the Bohm alternative formulation of
Quantum Mechanics. In this theory there is a hidden variable
namely the definite value of position while the so called Bohm
potential $Q$ can be non local, two features which
do not find favour with physicists.\\
It must be noted that in Weyl's geometry, even in a Euclidean
space there is a
covariant derivative and a non vanishing curvature $R$.\\
Santamato \cite{cu1,cu2,cu3} exploits this latter fact, within the
context of the deBroglie-Bohm theory and postulates a Lagrangian
given by
$$L (q,{\dot q} , t) = L_c(q, {\dot q} ,t) + \gamma
(\hbar^2/m)R(q,t),$$
He then goes on to obtain the equations of
motion like (\ref{e1a}),(\ref{e2a}), etc. by invoking an Averaged
Least Action Principle
$$I (t_0,t_1) = E \left\{ \int^t_{t_0} L^* (q(t,\omega),{\dot q} (t,\omega),t)dt\right\}$$
\begin{equation}
= \mbox{minimum} ,\label{e7a}
\end{equation}
with respect to the class of all Weyl geometries of space with
fixed metric tensor. This now leads to the Hamilton-Jacobi
equation
\begin{equation}
\partial_t S + H_c (q,\nabla S,t) - \gamma (\hbar^2 /m) R = 0,\label{e8a}
\end{equation}
and thence to the Schrodinger equation (in curvi-linear
coordinates)
$$\imath \hbar \partial_t \psi = (1/2m)\left\{ [(\imath \hbar /\sqrt{g})
\partial_\imath \sqrt{g} A_\imath ]g^{\imath k} (\imath \hbar \partial_k
+ A_k)\right\} \psi$$
\begin{equation}
+ [V - \gamma (\hbar^2 /m){\dot R} ] \psi = 0,\label{e9a}
\end{equation}
As can be seen from the above, the Quantum potential $Q$ is now
given in
terms of the scalar curvature $R$.\\
We have already related the arbitrary functions $\Phi$ of Weyl's
formulation with a noncommutative spacetime geometry (\ref{e7}).\\
This legitimises Santamato's postulative
approach of extending the deBroglie-Bohm formulation.\\
At an even more fundamental level, this formalism gives us the
rationale for the deBroglie wave length itself. Because of the
noncommutative geometry in (\ref{e7}) space becomes multiply
connected, in the sense that a closed circuit cannot be shrunk to
a point within the interval. Let us consider the simplest case of
double connectivity. In this case, if the interval is of length
$\lambda$, we will have,
\begin{equation}
\Gamma \equiv \int_c m\vec V \cdot d \vec r = h \int_c \vec \nabla
S \cdot d \vec r = h \oint dS = mV \pi \lambda = \pi h\label{e11a}
\end{equation}
whence
\begin{equation}
\lambda = \frac{h}{mV}\label{e12a}
\end{equation}
In (\ref{e11a}), the circuit integral was over a circle of
diameter $\lambda$. Equation (\ref{e12a}) shows the emergence of
the deBroglie wavelength. This follows from the noncommutative
geometry of space time, rather than the physical Heisenberg
Uncertainty Principle. Remembering that $\Gamma$ in (\ref{e11a})
stands for the angular momentum, this is also the origin of the
Wilson-Sommerfeld quantization rule, an otherwise mysterious
Quantum Mechanical prescription.\\
What we have done is to develop a Quantum Geometry, based on
(\ref{e7}).
\section{Cosmology}
In recent years the work of Perlmutter and co-workers has shown
that the universe is not only not descelerating, it is actually
accelerating, and would continue to expand for ever. The work of
Webb and co-workers on the other hand brings out an equally
iconoclastic observation: The hallowed fine structure constant is
actually slowly decreasing with time. Suddenly dark matter has
been discarded in favour of dark energy and a cosmological
constant.\\
We first observe that the concept of a Zero Point Field (ZPF) or
Quantum Vacuum (or Ether) is an idea whose origin can be traced
back to Max Planck himself. Quantum Field Theory attributes the
ZPF to the virtual Quantum Effects of an already present
electromagnetic field. There is another approach, sometimes called
Stochastic Electrodynamics which treats the ZPF as primary and
attributes to it Quantum Mechanical effects \cite{r12b,r13b}. It
may be observed that the ZPF results in the well known
experimentally verified Casimir effect \cite{r14b,r15b}. We would
also like to point out that contrary to popular belief, the
concept of Ether has survived over the decades through the works
of Dirac, Vigier, Prigogine, String Theorists like Wilzeck and
others \cite{r16b}-\cite{r24b}. It appears that even Einstein
himself
continued to believe in this concept \cite{r25b}.\\
We would first like to observe that the energy of the fluctuations
in the background electromagnetic field could lead to the
formation of elementary particles. Infact it is known that this
energy of fluctuation in a region of length $l$ is given by
\cite{MWT}
$$B^2 \sim \frac{\hbar c}{l^4}$$
In the above if $l$ is taken to be the Compton wavelength of a
typical elementary particle, then we recover its energy $mc^2$, as
can be easily verified. It may be mentioned that Einstein himself
had believed that the electron was a result of such condensation
from the background electromagnetic field (Cf.\cite{cu} for
details). Infact this formation of particles could be likened to
the formation of Benard cells in a phase transition \cite{r26b},
as we will see briefly in the next section. We also take the pion
to represent a typical elementary
particle, as in the literature.\\
To proceed, as there are $N \sim 10^{80}$ such particles in the
universe, we get
\begin{equation}
Nm = M\label{e1b}
\end{equation}
where $M$ is the mass of the universe.\\
In the following we will use $N$ as the sole cosmological parameter.\\
Equating the gravitational potential energy of the pion in a three
dimensional isotropic sphere of pions of radius $R$, the radius of
the universe, with the rest energy of the pion, we can deduce the
well known relation \cite{cu}
\begin{equation}
R \approx \frac{GM}{c^2}\label{e2b}
\end{equation}
where $M$ can be obtained from (\ref{e1b}).\\
We now use the fact that given $N$ particles, the fluctuation in
the particle number is of the order $\sqrt{N}$\cite{r29b}, while a
typical time interval for the fluctuations is $\sim \hbar/mc^2$,
the Compton time. We will come back to this point later. So we
have
$$\frac{dN}{dt} = \frac{\sqrt{N}}{\tau}$$
whence on integration we get,
\begin{equation}
T = \frac{\hbar}{mc^2} \sqrt{N}\label{e3b}
\end{equation}
We can easily verify that equation (\ref{e3b}) is indeed satisfied
where $T$ is the age of the universe. Next by differentiating
(\ref{e2b}) with respect to $t$ we get
\begin{equation}
\frac{dR}{dt} \approx HR\label{e4b}
\end{equation}
where $H$ in (\ref{e4b}) can be identified with the Hubble
Constant, and using (\ref{e2b}) is given by,
\begin{equation}
H = \frac{Gm^3c}{\hbar^2}\label{e5b}
\end{equation}
Equation (\ref{e1b}), (\ref{e2b}) and (\ref{e3b}) show that in
this formulation, the correct mass, radius and age of the universe
can be deduced given $N$ as the sole cosmological or large scale
parameter. Equation (\ref{e5b}) can be written as
\begin{equation}
m \approx
\left(\frac{H\hbar^2}{Gc}\right)^{\frac{1}{3}}\label{e6b}
\end{equation}
Equation (\ref{e6b}) has been empirically known as an "accidental"
or "mysterious" relation. As observed by Weinberg\cite{r30b}, this
is unexplained: it relates a single cosmological parameter $H$ to
constants from microphysics. We will touch upon this micro-macro
nexus again. In our formulation, equation (\ref{e6b}) is no longer
a mysterious coincidence but
rather a consequence.\\
As (\ref{e5b}) and (\ref{e4b}) are not exact equations but rather,
order of magnitude relations, it follows that a small cosmological
constant $\wedge$ is allowed such that
$$\wedge \leq 0 (H^2)$$
This is consistent with observation and shows that $\wedge$ is
very very small - this has been a puzzle, the so called
cosmological constant problem \cite{r31b}. But it is explained
here.\\
To proceed we observe that because of the fluctuation of $\sim
\sqrt{N}$ (due to the ZPF), there is an excess electrical
potential energy of the electron, which infact we have identified
as its inertial energy. That is \cite{r29b},
$$\sqrt{N} e^2/R \approx mc^2.$$
On using (\ref{e2b}) in the above, we recover the well known
Gravitation-electromagnetism ratio viz.,
\begin{equation}
e^2/Gm^2 \sim \sqrt{N} \approx 10^{40}\label{e7b}
\end{equation}
or without using (\ref{e2b}), we get, instead, the well known so
called Eddington formula,
\begin{equation}
R = \sqrt{N}l\label{e8b}
\end{equation}
Infact (\ref{e8b}) is the spatial counterpart of (\ref{e3b}). If
we combine (\ref{e8b}) and (\ref{e2b}), we get,
\begin{equation}
\frac{Gm}{lc^2} = \frac{1}{\sqrt{N}} \propto T^{-1}\label{e9b}
\end{equation}
where in (\ref{e9b}), we have used (\ref{e3b}). Following Dirac
(cf.also \cite{r32b}) we treat $G$ as the variable, rather than
the quantities $m, l, c \mbox{and} \hbar$ (which we will call
micro physical constants) because of their central role
in atomic (and sub atomic) physics.\\
Next if we use $G$ from (\ref{e9b}) in (\ref{e5b}), we can see
that
\begin{equation}
H = \frac{c}{l} \quad \frac{1}{\sqrt{N}}\label{e10b}
\end{equation}
Thus apart from the fact that $H$ has the same inverse time
dependance on $T$ as $G$, (\ref{e10b}) shows that given the
microphysical constants, and
$N$, we can deduce the Hubble Constant also, as from (\ref{e10b}) or (\ref{e5b}).\\
Using (\ref{e1b}) and (\ref{e2b}), we can now deduce that
\begin{equation}
\rho \approx \frac{m}{l^3} \quad \frac{1}{\sqrt{N}}\label{e11b}
\end{equation}
Next (\ref{e8b}) and (\ref{e3b}) give,
\begin{equation}
R = cT\label{e12b}
\end{equation}
(\ref{e11b}) and (\ref{e12b}) are consistent with observation.\\
The above model predicts an ever expanding and possibly
accelerating universe whose density keeps decreasing. This seemed
to go against the accepted idea that the density of the universe
equalled the critical density required for closure.\\
The above cosmology exhibits a time variation of the gravitational
constant of the form
\begin{equation}
G = \frac{\beta}{T}\label{e14b}
\end{equation}
Indeed this is true in a few other schemes also, including Dirac's
cosmology (Cf. \cite{r33b}). Interestingly it can be shown that
such a time variation can explain the precession of the perihelion
of Mercury (Cf.\cite{r35b}). It can also provide an alternative
explanation for dark matter and the bending of light while the
Cosmic Microwave Background
Radiation is also explained (Cf.\cite{cu}).\\
It is also possible to deduce the existence of gravitational waves
given (\ref{e14b}). To see this quickly let us consider the
Poisson equation for the metric $g_{\mu \nu}$
\begin{equation}
\nabla^2 g_{\mu \nu} = G \rho u_\mu u_\nu\label{e15b}
\end{equation}
The solution of (\ref{e15b}) is given by
\begin{equation}
g_{\mu \nu} = G \int \frac{\rho u_\mu u_\nu}{|\vec r - \vec r'|}
d^3\vec r\label{e16b}
\end{equation}
Indeed equations similar to (\ref{e15b}) and (\ref{e16b}) hold for
the Newtonian gravitational potential also. If we use the second
time derivative of $G$ from (\ref{e14b}) in (\ref{e16b}), along
with (\ref{e15b}), we can immediately obtain the D'alembertian
wave equation for gravitational waves, instead of the Poisson
equation:
$$D g_{\mu \nu} \approx 0$$
where $D$ is the D'alembertian.\\
Recently a small variation with time of the fine structure
constant has been detected and reconfirmed by Webb and coworkers
\cite{r36b,r37b}. This observation is consistent with the above
cosmology. We can see this as follows. We use an equation due to
Kuhne \cite{r38b}
\begin{equation}
\frac{\dot \alpha_z}{\alpha_z} = \alpha_z \frac{\dot
H_z}{H_z},\label{e17b}
\end{equation}
If we now use the fact that the cosmological constant $\Lambda$ is
given by
\begin{equation}
\Lambda \leq 0(H^2)\label{e18b}
\end{equation}
as can be seen from (\ref{e4b}), in (\ref{e17b}), we get using
(\ref{e18b}),
\begin{equation}
\frac{\dot \alpha_z}{\alpha_z} = \beta H_z\label{e19b}
\end{equation}
where $\beta < - \alpha_z < - 10^{-2}$.\\
Equation (\ref{e19b}) can be shown to be the same as
\begin{equation}
\frac{\dot \alpha_z}{\alpha_z} \approx -1 \times 10^{-5}
H_z.\label{e20b}
\end{equation}
which is the same as Webb's result.\\
We give another derivation of (\ref{e20b}) in the above context
wherein, as the number of particles in the universe increases with
time, we go from the Planck
scale to the Compton scale.\\
This can be seen as follows: In equation (\ref{e7b}), if the
number of particles in the universe, $N = 1$, then the mass $m$
would be the Planck mass. In this case the classical Schwarzschild
radius of the Planck mass would equal its Quantum Mechanical
Compton wavelength. To put it another way, all the energy would be
gravitational (Cf.\cite{cu} for details). However as the number of
particles $N$ increases with time, according to (\ref{e3b}),
gravitation and electromagnetism
get differentiated and we get (\ref{e7b}) and the Compton scale.\\
It is known that the Compton length, due to zitterbewegung causes
a correction to the electrostatic potential which an
orbiting electron experiences, rather like the Darwin term \cite{bd}.\\
Infact we have
$$\langle \delta V \rangle = \langle V (\vec r + \delta \vec r)\rangle - V
\langle (\vec r )\rangle$$
$$= \langle \delta r \frac{\partial V}{\partial r} + \frac{1}{2} \sum_{\imath j}
\delta r_\imath \delta r_j \frac{\partial^2 V}{\partial r_\imath
\partial r_j} \rangle$$
\begin{equation}
\approx 0(1) \delta r^2 \nabla^2 V\label{e21b}
\end{equation}
Remembering that $V = e^2/r$ where $r \sim 10^{-8}cm$, from
(\ref{e21b}) it follows that if $\delta r \sim l$, the Compton
wavelength then
\begin{equation}
\frac{\Delta \alpha}{\alpha} \sim 10^{-5}\label{e22b}
\end{equation}
where $\Delta \alpha$ is the change in the fine structure constant
from the early universe. (\ref{e22b}) is an equivalent form of
(\ref{e20b}) (Cf.ref.\cite{r38b,r39b}), and is the result
originally
obtained by Webb et al (Cf.refs.\cite{r39b,r7d}).\\
Infact there is another test for the variation of $G$.We would now
like to show that this model also explains the observed decrease
in the orbital period of the binary pulsar PSR $1913 + 16$,
otherwise
attributed to as yet undetected gravitational waves \cite{r7d}.\\
In general in schemes in which $G$ the universal constant of
gravitation decreases with time, it is to be expected that
gradually the size of the orbit and the time period would
increase, with an overall decrease in energy. In any case all this
becomes more relevant in the light of the above latest
observations about the fine structure constant.\\
But in the present case as we will show, the gravitational energy
of the binary system, $\frac{GMm}{L}$ remains constant, where $M$
is the mass of the central object and $L$ the mean distance
between the objects. This is because the decrease in $G$ is
compensated by an increase in the material content of the system,
according to the
above model.\\
In fact the energy lost is given by $\frac{GM}{TL}$ (per unit mass
of the orbiting object - in any case the mass of the orbiting
object does not feature in the dynamical equations). Further from
what we saw $\frac{1}{\sqrt{N}\tau} = \frac{1}{T}$ particles
appear from the Quantum Vacuum per second, per particle in the
universe. So the energy gained in this process is $\frac{GM}{TL}$
per second. This follows, if we write $M = n \times m$, where $n$
is the number of typical elementary particles in the central body
and $m$ their
mass.\\
As can be seen from the above the energy lost per second is
compensated by the energy
gained and thus the total gravitational energy of the binary system remains constant.\\
Let us now consider an object revolving about another object, as
in the case of the binary pulsar. The gravitational energy of the
system is now given by,
$$\frac{GMm}{L} = const.$$
Whence
\begin{equation}
\frac{\mu}{L} \equiv \frac{GM}{L} = const.\label{e3d}
\end{equation}
For variable $G$ we have
\begin{equation}
\mu = \mu_0 - tK\label{e4d}
\end{equation}
where
\begin{equation}
K \equiv \dot \mu\label{e5d}
\end{equation}
We take $\dot \mu$ to be a constant, in view of the fact that $G$
varies very slowly, as can be seen from (\ref{e1}).\\
To preserve (\ref{e3d}), we should have
$$L = L_0 (1 - \alpha K)$$
Whence on using (\ref{e4d})
\begin{equation}
\alpha = \frac{t}{\mu_0}\label{e6d}
\end{equation}
We shall consider $t$, to be the period of revolution. Using
(\ref{e6d}) it follows that
\begin{equation}
\delta L = - \frac{LtK}{\mu_0}\label{e7d}
\end{equation}
We also know from theory,
\begin{equation}
t = \frac{2 \pi}{h} L^2 = \frac{2\pi}{\sqrt{\mu}}\label{e8d}
\end{equation}
\begin{equation}
t^2 = \frac{4\pi^2 L^3}{\mu}\label{e9d}
\end{equation}
Using (\ref{e7d}), (\ref{e8d}) and (\ref{e9d}), a little
manipulation gives
\begin{equation}
\delta t = - \frac{2t^2K}{\mu_0}\label{e10d}
\end{equation}
(\ref{e7d}) and (\ref{e10d}) show that there is a decrease in the
size of the orbit, as also in the orbital period. Before
proceeding further we note that such a decrease in the orbital
period has been observed in the case of binary
pulsars \cite{r7d,r13d}.\\
Let us now apply the above considerations to the case of the
binary pulsar PSR $1913 + 16$ observed by Taylor and co-workers
(Cf.ref.\cite{r13d}). In this case it is known that, $t$ is 8
hours while $v$, the orbital speed is $3 \times 10^7cms$ per
second. It is easy to calculate from the above
$$\mu_0 = 10^4 \times v^3 \sim 10^{26}$$
which gives $M \sim 10^{33}gms$, which of course agrees with
observation. Further we get using (\ref{e5d})
\begin{equation}
\Delta t = \eta \times 10^{-5}sec/yr, \eta \leq 8\label{e11d}
\end{equation}
Indeed (\ref{e11d}) is in good agreement with the carefully
observed
value of $\eta \approx 7.5$ (Cf.refs.\cite{r7d,r13d}).\\
It may be remarked that this same effect has been interpreted as
being due to gravitational radiation, even though there are some
objections to the calculation in this case (Cf.ref.\cite{r13d}).\\
We will now try to explain the Pioneer spacecrafts' large
anomalous acceleration which has been studied for several years by
Anderson and co-workers and has remained a long standing puzzle
\cite{r14d}.\\
From the energy conservation in central orbits, viz.,
$$\frac{m}{2} (\dot r^2 + r^2 \dot \Theta^2) - \frac{GM}{r} =
const.,$$
we get, on differentiation, using the effect of the
variable $G$ (\ref{e9b}), an extra inward acceleration,
\begin{equation}
a_r = \frac{GM}{t_0r\dot r}\label{eA}
\end{equation}
where $t_0$ is the age of the universe.\\
On the other hand, from the standard equation for the orbit
$$\frac{l}{r} = 1+e cos \Theta ,$$
where $l = \frac{(r^2 \dot \Theta)^2}{GM}$, we get,
differentiating and using (\ref{e9b}), the extra effect,
\begin{equation}
\dot r \approx \frac{r^{3/4}\nu}{t_0 \sqrt{GM}},\label{eB}
\end{equation}
where $\nu = r \dot \Theta$.\\
Using (\ref{eB}) in (\ref{eA}) and the values for $r$ and $\nu$,
viz., $\sim 10^{15}$ and $\sim 10^6$ respectively, we get
$$a_r \leq 10^{-6} cm/sec^2$$
This fits in very well with the results of Anderson et al that
$$a_r \sim 10^{-7} cm/sec^2$$
Thus, we can argue that the inexplicable large anomalous
acceleration is a footprint of the variable $G$ (\ref{e9b}).
\section{Critical Phenomena and Cosmology}
It has been pointed out that in the universe at large, there
appears to be the analogues of the Planck constant $h$ at
different scales \cite{r1,r2,cu} and several references therein.
Infact we have
\begin{equation}
h_1 \sim 10^{93}\label{e1c}
\end{equation}
for super clusters;
\begin{equation}
h_2 \sim 10^{74}\label{e2c}
\end{equation}
for galaxies and
\begin{equation}
h_3 \sim 10^{54}\label{e3c}
\end{equation}
for stars. And
\begin{equation}
h_4 \sim 10^{34}\label{e4c}
\end{equation}
for Kuiper Belt objects. In equations (\ref{e1c}) - (\ref{e4c}),
the $h_\imath$ play the role of the Planck constant, in a sense to
be described below. The origin of these equations is related to
the following empirical relations
\begin{equation}
R \approx l_1 \sqrt{N_1}\label{e5c}
\end{equation}
\begin{equation}
R \approx l_2 \sqrt{N_2}\label{e6c}
\end{equation}
\begin{equation}
l_2 \approx l_3 \sqrt{N_3}\label{e7c}
\end{equation}
\begin{equation}
R \sim l \sqrt{N}\label{e8c}
\end{equation}
and a similar relation for the KBO (Kuiper Belt objects)
\begin{equation}
L \sim l_4 \sqrt{N_4}\label{e65}
\end{equation}
where $N_1 \sim 10^6$ is the number of superclusters in the
universe, $l_1 \sim 10^{25}cms$ is a typical supercluster size
$N_2 \sim 10^{11}$ is the number of galaxies in the universe and
$l_2 \sim 10^{23}cms$ is the typical size of a galaxy, $l_3 \sim
1$ light years is a typical distance between stars and $N_3 \sim
10^{11}$ is the number of stars in a galaxy, $R$ being the radius
of the universe $\sim 10^{28}cms, N \sim 10^{80}$ is the number of
elementary particles, typically pions in the universe and $l$ is
the pion Compton wavelength and $N_4 \sim 10^{10}, l_4 \sim 10^5
cm$, is the age of a typical KBO (with mass $10^{19}gm$ and $L$
the width of the Kuiper Belt $\sim 10^{10}cm$
cf.ref.\cite{cu}).\\
The size of the universe is the size of a supercluster etc. from
equations like (\ref{e5c})-(\ref{e8c}), as described in the
references turn up as the analogues of the Compton wavelength. For
example we have
\begin{equation}
R = \frac{h_1}{Mc}\label{e9c}
\end{equation}
It has also been argued that these scaled Compton wavelengths and
scaled Planck constants are the result of gravitational orbits
described by
\begin{equation}
\frac{GM}{L} \sim v^2\label{e10c}
\end{equation}
For example from (\ref{e10c}) it follows that
$$M v L = h_2,$$
Whence we get (\ref{e2c}).\\
While equations (\ref{e5c})-(\ref{e8c}), resemble the Random Walk relations, this is not
accidental.\\
Let us start with Nelson's well known equation of diffusion
\begin{equation}
\Delta k^2 = \nu \Delta t, \quad \nu = \frac{h}{m}\label{e11c}
\end{equation}
where $\nu$ is the diffusion constant, $h$ the Planck constant and
$m$ the mass of a typical elementary particle like pion. We can
see immediately from (\ref{e11c}) that for velocities
$$\langle \frac{\Delta x}{\Delta t} \rangle = c$$
We recover the Compton wavelength (or more generally the deBroglie
wavelength). Further it can be shown that from (\ref{e11c}) we can
recover the Random Walk equation (\ref{e8c}) \cite{r4} define
scaled diffusion constant to be $\nu_\imath$. Using (\ref{e11c})
it immediately follows that
\begin{equation}
\Delta x^2_\imath = \nu_\imath \Delta t_\imath\label{e12c}
\end{equation}
(\ref{e12c}) is a diffusion equation at different scales. This in
turn leads to equations (\ref{e5c}) - (\ref{e65}), a similar one
for the KBO, which were to start with empirical relations. The
interesting point is that starting from (\ref{e11c}) (or
(\ref{e12c})), we can infact deduce the Schrodinger equation -
either by using the Lagrangian, as is known, or even without
invoking a Lagrangian (Cf.Appendix):
\begin{equation}
h_\imath \frac{\partial \psi}{\partial t} + \frac{h^2_\imath}{2m}
\nabla^2 \psi = 0\label{e13c}
\end{equation}
This provides a rationale for the scaled Compton lengths or scaled deBroglie lenghs
referred to above.\\
Let us investigate the above considerations in a little more detail.\\
For this, we observe that the creation of particles from a Quantum
vacuum (or pre space time) has been described in the references
cited \cite{r5a,r6}. It can be done within the context of the
above Nelsonian Theory, in complete analogy with the creation of
Benard cells at the critical point. The Nelsonian-Brownian process
as described in (\ref{e11c}) defines, first the Planck length, the
shortest possible length and then the random process leads to the
Compton scale (Cf.ref.\cite{r4}). This process is as noted
(Cf.ref.\cite{r6}) a complete analogue of the phase transition
associated with the Landau-Ginsburg equation \cite{r7}
\begin{equation}
-\frac{h^2}{2m} \nabla^2 \psi + \beta |\psi |^2 \psi = -\propto
\psi\label{e14c}
\end{equation}
The parallel is not yet fully apparent, if we compare (\ref{e14c})
and the Schrodinger equation (\ref{e13c}). However this becomes
clear if we consider how the Schrodinger equation itself can be
deduced from the amplitudes of the Quantum vacuum, in which case
we get
\begin{equation}
\imath \hbar \frac{\partial \psi}{\partial t} =
\frac{-\hbar^2}{2m'}\frac{\partial^2 \psi}{\partial x^2} + \int
\psi^* (x')\psi (x)\psi (x')U(x')dx',\label{e15c}
\end{equation}
infact the correlation length from (\ref{e14c}) is given by
$$\xi = (\frac{\gamma}{\propto})^{\frac{1}{2}}$$
which can be easily reduced to the Compton wavelength. In other
words, the Schrodinger equation (\ref{e13c}), via (\ref{e15c})
describes the creation of particles,
a la Benard cells in a Landau-Ginsburg like phase transition.\\
As is known, the interesting aspects of the critical point theory
(Cf.ref.\cite{r7,r8a}) are universality and scale. Broadly, this
means that diverse physical phenomena follow the same route at the
critical point, on the one hand, and on the other this can happen
at different scales, as exemplified for example, by the course
graining techniques of the renormalization group. To highlight
this point we note that in critical point phenomena we have the
reduced order parameter $\bar Q$ and the reduced correlation
length bar $\bar \xi$. Near the critical point we have relations
like (Cf.ref.\cite{r8a})
$$(\bar Q) = |t|^\beta , (\bar \xi) = |t|^{-\nu}$$
Whence
\begin{equation}
\bar Q^\nu = \bar \xi^\beta\label{e16c}
\end{equation}
In (\ref{e16c}) typically $\nu \approx 2\beta$. As $\sqrt{Q} \sim
\frac{1}{\sqrt{N}}$ because $\sqrt{N}$ particles are created
fluctuationally, given $N$ particles, and in view of the fractal
two dimensionality of the path
$$\bar Q \sim \frac{1}{\sqrt{N}}, \bar \xi = (l/R)^2$$
This gives
$$R = \sqrt{N}l$$
which is nothing but (\ref{e8c}).\\
In other words the scaled Planck effects and the scaled Random
Walk effects as typified by equations like (\ref{e1c})-(\ref{e65})
are the result of a critical point phase transition and subsequent
course graining.
\section{Discussion}
1.In some ways the General Relativistic gravitational field
resembles the electromagnetic field, particularly in certain
approximations, as for example when the field is stationary or
nearly so and the velocities are small. In this case the equations
of General Relativity can be put into a form resembling those of
Maxwell's Theory, and then the fields have been called
Gravitoelectric and Gravitomagnetic \cite{r1e}. Experiments have
also been suggested for measuring the Gravitomagnetic
force components for the earth \cite{r3e}.\\
We can ask whether such a consideration can be applied to
elementary particles, if in fact they can be considered in the
context of General Relativity. It may be mentioned that apart from
Quantum Gravity, there have been three different approaches for
studying elementary particles via General Relativity
\cite{r4e,r5e,cu} and references therein. We will now show that it
is possible to extend the Gravitomagnetic and Gravitoelectric
formulations to elementary particles within the framework of the
theory developed in \cite{cu}.\\
In \cite{cu}, the linearized General Relativistic equations are
seen to describe the properties of elementary particles, such as
spin, mass, charge and even the very Quantum Mechanical anomalous
gyromagnetic ratio $g = 2$, apart from several other
characteristics \cite{r7e,r8e,r9e,r10e}.\\
We merely report that the linearized equations of General
Relativity, viz.,
\begin{equation}
g_{\mu \nu} = \eta_{\mu \nu} + h_{\mu \nu}, h_{\mu \nu} = \int
\frac{4T_{\mu \nu}(t-|\vec x - \vec x'|, \vec x')}{|\vec x - \vec
x'|} d^3 x'\label{e1da}
\end{equation}
where as usual,
\begin{equation}
T^{\mu \nu} = \rho u^u u^v\label{e2da}
\end{equation}
lead to the mass, spin, gravitational potential and charge of an
electron, if we work at the Compton scale (Cf.ref.\cite{cu} for
details). Let us now apply the macro Gravitoelectic and
Gravitomagnetic equations to the above case. Infact these
equations are (Cf.ref.\cite{r1e}).
\begin{equation}
\nabla \cdot \vec E_g \approx -4\pi \rho, \nabla \times \vec E_g
\approx - \partial \vec H_g/\partial t, etc.\label{e3da}
\end{equation}
\begin{equation}
\vec E_g = - \nabla \phi - \partial \vec A/\partial t, \quad \vec
H_g = \nabla \times \vec A\label{e4da}
\end{equation}
\begin{equation}
\phi \approx - \frac{1}{2} (g_{00} + 1), \vec A_\imath \approx
g_{0 \imath},\label{e5da}
\end{equation}
The subscripts $g$ in the equations (\ref{e3da}), (\ref{e4da}),
(\ref{e5da}) are to indicate that the fields $E$ and $H$ in the
macro case do not really represent the Electromagnetic field, but
rather resemble them. Let us apply equation (\ref{e4da}) to
equation (\ref{e1da}), keeping in mind equation (\ref{e5da}). We
then get, considering only the order of magnitude, which is what
interests us here, after some manipulation
\begin{equation}
|\vec H | \approx \int \frac{\rho V}{r^2} \bar r \approx \frac{m
V}{r^2}\label{e6da}
\end{equation}
and
\begin{equation}
| \vec E | = \frac{mV^2}{r^2}\label{e7da}
\end{equation}
$V$ being the speed.\\
In (\ref{e6da}) and (\ref{e7da}) the distance $r$ is much greater
than a typical Compton wavelength, to make the approximations
considered in
deriving the Gravitomagnetic and Gravitoelectric equations meaningful.\\
Remembering that we have
$$mVr \approx h,$$
the electric and magnetic fields in (\ref{e6da}) and (\ref{e7da})
now become
\begin{equation}
|\vec H | \sim \frac{h}{r^3} , |\vec E | \sim
\frac{hV}{r^3}\label{e8da}
\end{equation}
We now observe that (\ref{e8da}) does not really contain the mass
of the
elementary particle. Could we get a further insight into this new force?\\
Indeed in the above linearized General Relativistic
characterisation of the electron, it turns out that the electron
can be represented by the Kerr-Newman metric (Cf.\cite{cu} for
details). This incidentally also gives the anomalous gyromatgnetic
ratio $g=2$. This result has recently been reconfirmed by Nottale
\cite{r12e} from a totally different point of view, using scaled
relativity. It is well known that the Kerr-Newman field has extra
electric and magnetic terms (Cf.\cite{r13e}),
both of the order $\frac{1}{r^3}$, exactly as indicated in (\ref{e8da}).\\
It may be asked if there is any candidate as yet for the above
short range force. There is already one such candidate - the
inexplicable $B_{(3)}$ \cite{r14e} force mediated by massive
photons and of short range, first detected in 1992 at Cornell and
since confirmed by subsequent experiments. (It differs from the
usual $B_{(1)}$ and $B_{(2)}$ fields of special relativity,
mediated as
they are, by massless photons.)\\
Interestingly, if we work with a massive vector field we can
recover (\ref{e7da}) and (\ref{e8da})\cite{iz}. In this case there
is an upper limit on the mass of the photon $\sim
10^{-48}g$.\\
A Final Comment: It is quite remarkable that equations like
(\ref{e3da}), (\ref{e4da}) and (\ref{e5da}) which resemble the
equations of electromagnetism, have in the usual macro
considerations no connection whatsoever with electromagnetism
except in appearance. This would seem to be a rather miraculous
coincidence. In fact the above considerations of section 2 and
linearized General Relativistic theory of the electron as also the
Kerr-Newman metric formulation, demonstrate that the resemblence
to electromagnetism is not an accident, because in this latter
formulation, both electromagnetism and gravitation arise from the
metric (Cf.also refs.\cite{r15e,r7e,r8e,cu}).\\
2. Newman deduced the now famous Kerr-Newman metric alluded to
nearly 40 years ago \cite{r2f,r3f}. There were two inexplicable
features. The first was, the use of complex
coordinates, and why such coordinates somehow represent spin.\\
The second puzzling feature was, why should this General
Relativistic metric so closely describe the anomalous gyro
magnetic ratio, $g = 2$ \cite{r4f,MWT}.\\
The difficulty is best brought out by the fact that the
Kerr-Newman metric when applied to the electron throws up a naked
singularity, when the distances become complex which is a reversal
of the above situation, where
complex coordinates were introduced in the first place.\\
We will now first see, why complex coordinates arise at all, and
what the ramifications are: In the process we will get the answers
to the above two puzzling questions.\\
To elaborate the above considerations, the horizon of the
Kerr-Newman metric becomes, for the electron complex:
\begin{equation}
r = \frac{GM}{c^2} + \imath b, b \equiv \left( \frac{G^2Q^2}{c^8}
+ a^2 - \frac{G^2M^2}{c^4}\right)^{1/2}\label{e1f}
\end{equation}
However it should be noticed that the coordinates for a Dirac
particle is given by
\begin{equation}
x = (c^2p_1 H^{-1}t) + \frac{\imath}{2} c\hbar (\alpha_1 - cp_1
H^{-1}) H^{-1}\label{e2f}
\end{equation}
Interestingly the imaginary terms in both (\ref{e1f}) and
(\ref{e2f}) are of the same order, namely the Compton wavelength,
$\frac{\hbar}{mc}$ of the electron. For the complex or
non-Hermitian coordinate in (\ref{e2f}) Dirac had argued
\cite{r6f,r7f} that our measurements of space time intervals are
imprecise and infact averaged over the order of the Compton scale,
and once such averages are taken, the imaginary or zitterbewegung
term disappears, and we return to
real or Hermitian coordinates.\\
Infact in Dirac's theory the operator $d_x \equiv \frac{d}{dx}$ is
a purely imaginary operator, and is given by
$$\delta x (d_x + \bar d_x) = \delta x^2 d_x \bar d_x = 0$$
if
$$0 (\delta x^2) = 0$$
as is tacitly assumed. However if
\begin{equation}
0 (\delta x^2) \ne 0\label{e3f}
\end{equation}
then the operator $d_x$ becomes complex, and therefore, also the
momentum operator, $p_x \equiv \imath \hbar d_x$ and the position
operator. In other words if (\ref{e3f}) holds good then we have to
deal with complex or non-Hermitian coordinates. The implication of
this is that (Cf.\cite{r7e} for details) space time becomes non-
commutative as we saw in (\ref{e7}).\\
We also saw that this leads directly to the Dirac equation at the
Compton scale.\\
In any case here is the mysterious origin of the complex
coordinates and spin. The complex coordinates lead to the
Kerr-Newman metric and the electron's field including the
anomalous gyro magnetic ratio which are symptomatic of the
electron's spin. It also means that the naked singularity is
shielded by the fuzzy spacetime (Dirac's original averages over
the zitterbewegung interval or equivalently the noncommutative
geometry (\ref{e7}).\\
i) It may be noticed that while the original General Relativistic
and Kerr- Newman formulations were for the macro universe, we on
the other hand
have applied it to the micro world.\\
ii). Once complex coordinates are introduced, as noted by Newman
there is a change of character \cite{r4f}: "Notice that the
magnetic moment $\mu = ea$ can be thought of as the imaginary part
of the charge times the displacement of the charge into the
complex region.... We can think of the source as having a complex
center of charge and that the magnetic moment is the moment of
charge about the center of charge....In other words the total
complex angular momentum vanishes around any point $z^a$ on the
complex world-line. From this complex point of view the spin
angular momentum is identical to orbital, arising from an
imaginary shift of origin rather than a real one... If one again
considers the particle to be "localized" in the sense that the
complex center of charge coincides with the complex center of
mass, one again obtains the Dirac gyromagnetic ratio..."\\
Infact this above complexification of coordinates has also been
worked out by Kaiser \cite{r16f,r17f}, but for the Dirac electron.
Kaiser found that such a complexification eliminates the
unphysical zitterbewegung. Dirac himself had noted that
zitterbewegung was a manifestation of the fact that while in a
physical sense our space time intervals cannot be made arbitrarily
small, in theory we consider point intervals, so that as noted
above, once an averaging over
the Compton scale is performed, zitterbewegung disappears.\\
What we would like to stress here is that, the complexification
carried out by Newman or Kaiser were mathematical devices - the
physical motivation or ramification was unclear. We argue that
complexification  is symptomatic of the fuzzy nature of spacetime
at the micro scale.\\
iii). Interestingly the complexification of coordinates can also
be related to an underpinning of a Nelsonian-stochastic process
\cite{r18f}.\\
3. Interestingly we can purse the reasoning of section 4,
equations (57) ff to the case of terrestrial phenomena. Let us
consider a gas at standard temperature and pressure. In this case,
the number of molecules $n \sim 10^{23}$ per cubic centimeter, so
that $r \sim 1 cm$ and with the same $l$, we can get a "scaled"
Planck
constant $\tilde h \sim 10^{-44} << h$, the Planck constant.\\
In this case, a simple application of the WKB approximation, leads
immediately from the Schrodinger equation at the new scale to the
classical Hamilton-Jacobi theory, that is to classical mechanics.

{\large APPENDIX}\\ \\
As mentioned  on using the double Weiner process and Newtonian
mechanics, it is possible to derive the Schrodinger equation -
this is the derivation of Quantum Mechanics from the Nelsonian
stochastic theory. Let us briefly review the steps in this
derivation. We first start with the backward and forward time
derivatives in the double Weiner process,
$$\partial \rho/\partial t + div(\rho b_+) = \nu \Delta \rho,$$
\begin{equation}
\partial \rho/\partial t + div(\rho b_-) = - \nu \Delta
\rho\label{e1g}
\end{equation}
Next we use the fact that if $\rho (\vec r,t)$ is the probability
density at $\vec r (t)$, then as demonstrated by Kolmogorov, for
the above Weiner process and more generally any Markov process, we
have forward and backward Fokker-Planck equations
\begin{equation}
\frac{d_+}{dt} x(t) = b_+ \quad , \quad \frac{d_-}{dt} x(t) =
b_-\label{e2g}
\end{equation}
where, in the Nelsonian theory $\nu = \hbar/2m, \hbar$ being the
reduced Planck constant. It is now possible from (\ref{e2}) to
define the velocities
\begin{equation}
V = \frac{b_++b_-}{2} \quad ; \quad U =
\frac{b_+-b_-}{2}\label{e3g}
\end{equation}
Adding both the equations of (\ref{e1g}) we get on using
(\ref{e3g}) the equation of continuity,
\begin{equation}
\partial \rho/\partial t + div(\rho V) = 0\label{e4g}
\end{equation}
while on subtracting the two equations of (\ref{e1g}) we get
\begin{equation}
U = \nu \nabla ln\rho\label{e5g}
\end{equation}
In the usual approach we define the complex velocity
$$\vee = V - \imath U$$
and on using a suitable Lagrangian derived from Newtonian
Mechanics we deduce the  Schrodinger  equation for details). In
any case it is to be noted that if the velocity $U$ as given in
(\ref{e3g}) or (\ref{e5g}) vanishes, that is the backward and
forward time derivatives become equal, there is no double Weiner
process and we have the usual Classical Theory, with $\nu = 0$. So
Quantum
Mechanics is contained in $U$).\\
Let us now not take recourse to Newtonian Mechanics, but merely
define a function $S$ such that
\begin{equation}
V = \nu \vec \nabla S\label{e6g}
\end{equation}
Using equations (\ref{e5g}) and (\ref{e6g}) it is possible to
define a complex velocity potential $\psi$ given by
\begin{equation}
\psi = \sqrt{\rho} e^{(\imath /\hbar)}{^S}\label{e7g}
\end{equation}
We get the same $\psi$ in the usual stochastic theory (as also in
the hydrodynamical formulation), as can be easily verified. We now
observe that substitution of $V$ in terms of $\psi$ from
(\ref{e7g}) in the equation of continuity (\ref{e4g}) immediately
leads to
\begin{equation}
\imath \hbar \frac{\partial \psi}{\partial t} + \frac{\hbar^2}{2m}
\nabla^2 \psi = 0\label{e8g}
\end{equation}
(\ref{e8g}) can be seen to be the free particle Schrodinger
equation. We have not however used Newtonian Mechanics in the
derivation of (\ref{e8g}). In the usual Nelsonian
Theory also, (\ref{e8g}) is deduced, though via a Lagrangian.\\
If we now specialise to energy momentum eigen states, in the
Quantum Mechanical Schrodinger equation (\ref{e8g}), we obtain
\begin{equation}
E = p^2/2m\label{e9g}
\end{equation}
which expresses Newtonian Mechanics, as obtained from stochastic
considerations. (For example a time derivative of both sides of
(\ref{e9g}) and a little algebraic manipulation, leads to Newton's
first and second laws.)
\end{document}